# An octahedral deformation with six α particles at the Z = 12 system, Mg nuclides: Third nucleons, Alpharons

Chang-Bum Moon*

*Hoseo University, Chung-Nam 336-795, Korea*

We suggest that the emergence of a large deformation in the magnesium, Mg, nuclides, especially at the Z = 12, N = 12, should be associated with an octahedral deformed shape. Within the framework of molecular geometrical symmetry, we find a possibility that the Z = 12, N = 12 system would form an octahedral structure consisting of six points of α($^4$He) particles, yielding the ground collectivity. With this point of view, we draw the following serial molecular structures; the Z = 10, N = 10, $^{20}$Ne, corresponds to a hexahedral, the Z = 8, N = 8, $^{16}$O, does to a tetrahedral, and the Z = 6, N = 6, $^{12}$C, does to a trigonal symmetry. Moreover, the Z = 2, N = 2, $^4$He(α), fits into a tetrahedral symmetry with four points of nucleons; two protons and two neutrons. The enhanced deformation at Z = 12 with N > 20 would be explained by a deformed shape related to an Ethene(Ethylene)-like skeleton with six α particles. The deformation at Z = 10, with N = 10 and 12, can be interpreted as being attributed to a hexahedral shape combined by five α particles as well. By noticing that the Z = 4, N = 4 system is unstable to the ground state under two-body system with two α particles, we conclude that α particles, rather than the eight-protons-neutrons nucleon, govern the $^8$Be stability. Accordingly, the α particle should be a third nucleon, like a proton or a neutron, in a nucleus. We name it the '***Alpharon***'. With this picture, we are able to open a new gate toward understanding of nuclear many-body systems; nucleon-nucleon interaction, shell structures, nucleon-synthesis, and nuclear matters. We argue that *nature favors three-body systems; three quarks for a nucleon, three nucleons for a nucleus*.

*cbmoon@hoseo.edu

## 1. Introduction

In molecular quantum world, there have been appeared fruitful of particular geometric shapes, leading to the symmetry properties of molecules themselves. It is essential to know how to classify any molecule according to its geometrical symmetry and how to use this classification for understanding the molecular properties [1]. The molecular symmetry properties can be described systematically by a group theory, in a sense, much of which is a summary of common sense about the symmetries of objects.

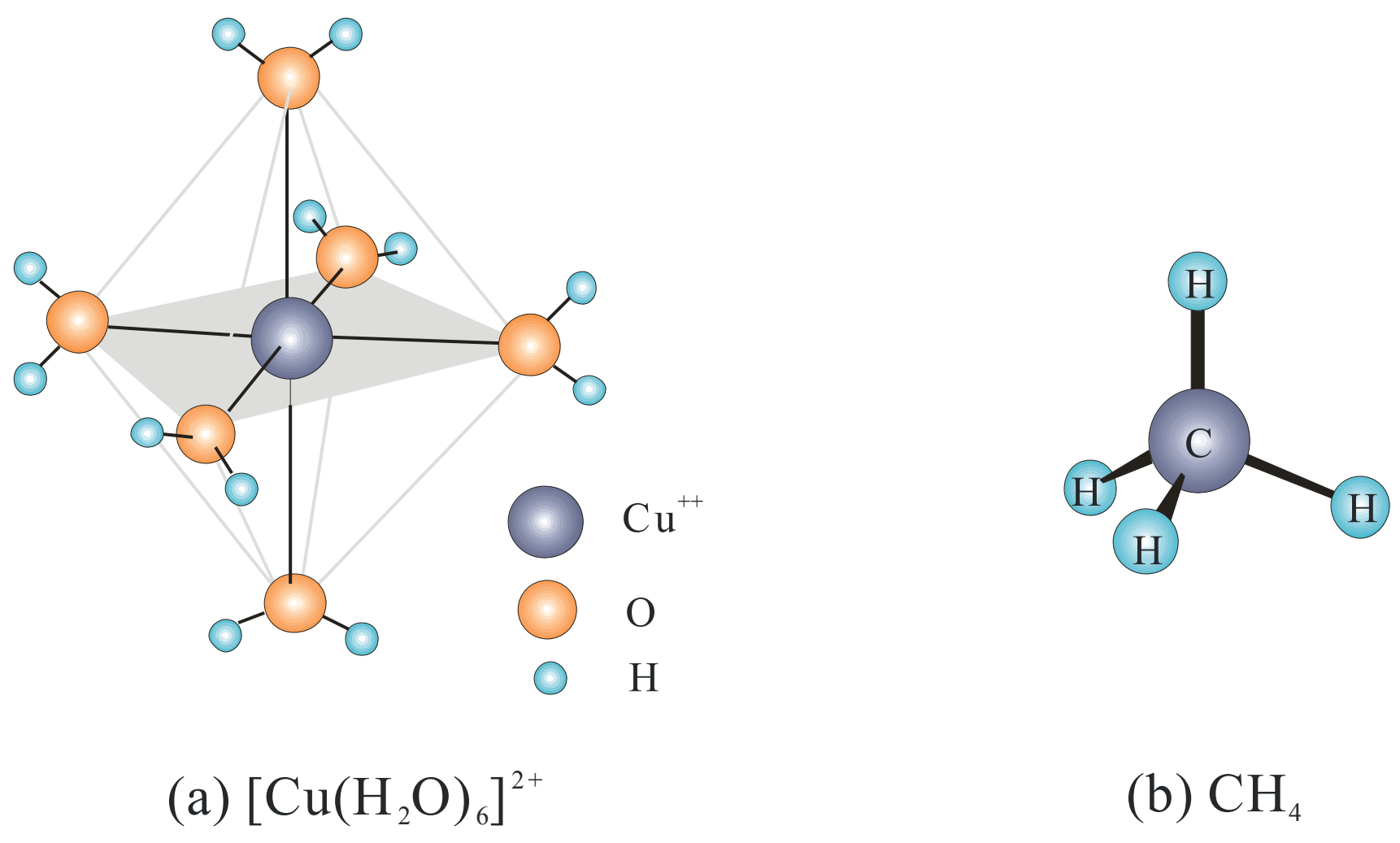

Fig. 1. Plots of geometrical shape for (a) the $[Cu(H_2O)_6]^{2+}$ complex, and (b) the methane, $CH_4$, molecules.

Figure 1 illustrates a geometrical picture of the molecule $[Cu(H_2O)_6]^{2+}$ complex and the methane molecule, where the octahedral and tetrahedral symmetries are seen. For the $[Cu(H_2O)_6]^{2+}$ molecule, the octahedron consists of six $H_2O$ ligands centered around $Cu^{++}$ ions. In the case of solid-state, the octahedron of the $La_2CuO_4$ compounds are also contributed from the Cu-O plane [2]. In materials science, the octahedral or tetrahedral structures play an important role in controlling specific properties such as super-conductivity, energy transfer, and fluorescence [2, 3].

Turning our attention to *atomic nuclear* physics, we find such an approach with the group theory associated with tetrahedral and octahedral symmetries for understanding the nuclear collective properties such as $^{156}$Gd and $^{160}$Yb [4, 5]. We argue that the very deformation at the Z = 12, N = 12 system, $^{24}$Mg, would come from an octahedral molecular structure combined by six α particles. In addition, we suggest that the $^{20}$Ne, $^{16}$O, and $^{12}$C nuclei might be interpreted as hexahedral, tetrahedral, and trigonal symmetries. In this work, any detailed descriptions with mathematical models, including quantitative results, are not provided. Instead, based on molecular geometrical symmetries, we offer the hypothetical models for describing nuclear structures in order to provide a new concept of nuclear physics theory. We demonstrate that α-decay processes from the Z = 12 to Z = 6 are explained in terms of serial geometrical symmetry breakings in our hypothetical model circumstances. On the basis of our molecular model descriptions, by interpreting some distinctive features in the α nuclides; $^{6}$He, $^{8}$He, and $^{10}$He, we suggest that α particles should be a member of nucleons in a nucleus, like protons or neutrons. Finally, we suppose that shape coexistence related to collectivity would appear in the Mg nuclides, which are associated with an octahedral and an Ethene-like structure consisting of six α nucleons, the so-called Alpharons. We show that the nucleus $^{3}$He should be an open system in a nucleus.

## 2. Systematic behaviors of the Mg nuclides

For describing fundamental characteristics of ground level properties in the nuclides, we begin by mapping the first $2^+$ excited energies and the ratios of the first $2^+$ and $4^+$ energies with respect to proton numbers, Z, and neutron numbers, N. See various systematic features based on these parameters in Refs. [8-11]. Figure 1 demonstrates systematics for those physical parameters within the space, Z = 8 to 28 [11].

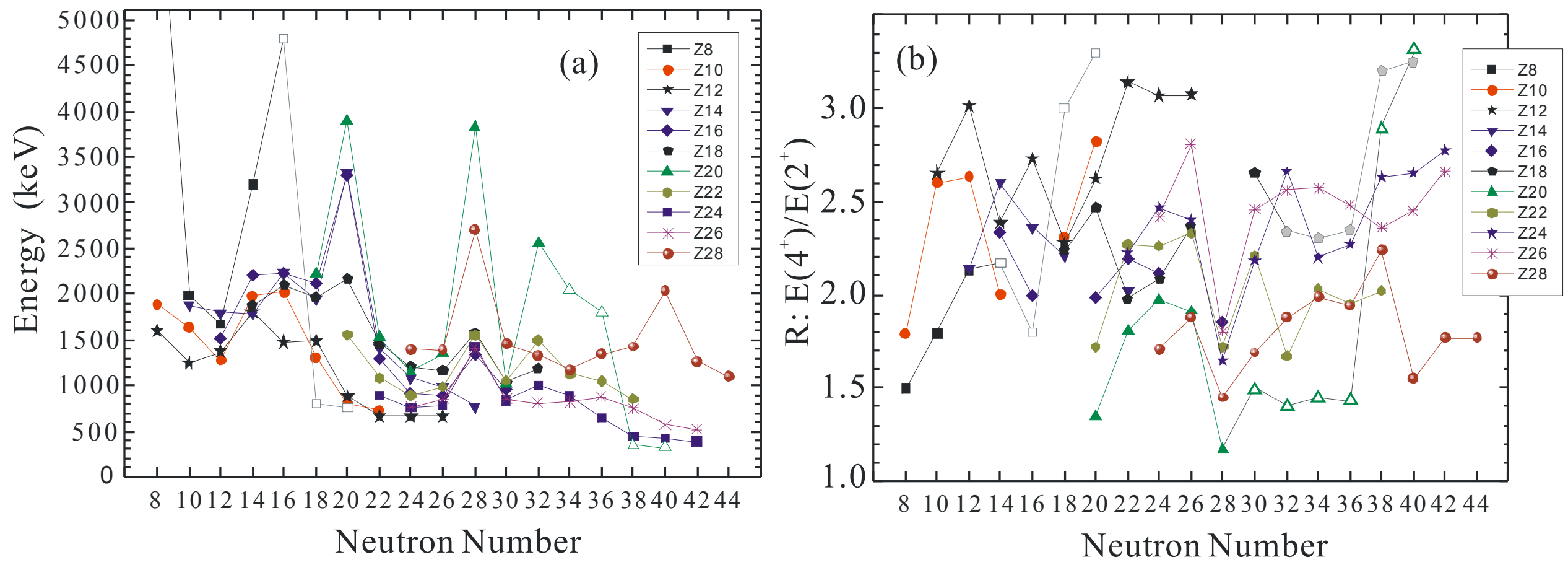

Fig. 2. (a) Systematic plots of $E(2^+)$ and (b) R, $E(4^+)/E(2^+)$, values as a function of neutron numbers for the nuclei within the Z = 8 and 28 space. Data are primarily taken from NNDC [6]. The points at Z = 12, N = 22, 24, 26 are from [7]. For the cases of Z = 8, 18, and 20, see Refs. [10, 11].

In the figure, we concentrate on some distinctive features for the Z = 12 regime; the $E(2^+)$ value at N = 16 and the R values at N = 12 and N = 16. This characteristic is neither seen at the Z = 10 nor at the Z = 14. For more visualizing characteristics related to the Mg nuclides, we demonstrate systematics of the low lying energy levels in Fig. 3, where the $0^+$, $2^+$, $4^+$, $1^-$, and $3^-$ states in Mg isotopes, including the R values within N = 8 to 26 are shown.

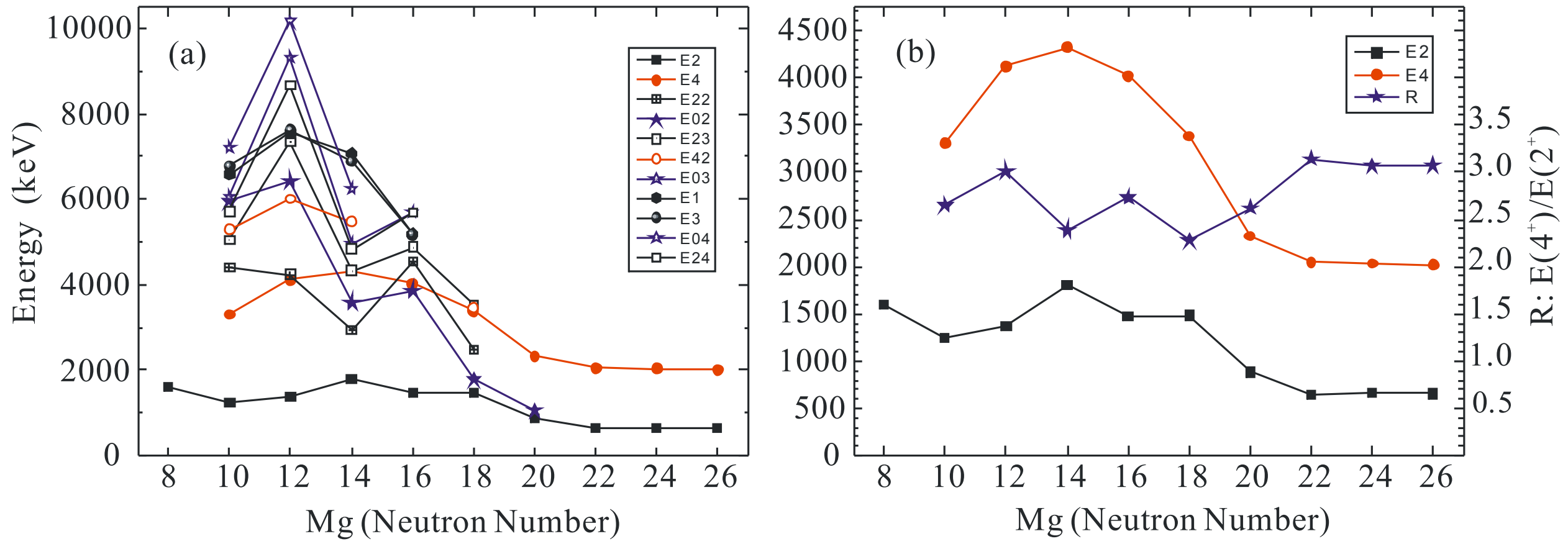

Fig. 3. (a) Systematic plots of some low-lying states in the Mg nuclides. In (b) R, $E(4^+)/E(2^+)$, values are plotted along with the $2^+$ and $4^+$ energies for comparisons.

For discussions, the deformation parameter, R value, is also included. Firstly, let us remark that the R values fluctuate with respect to neutron numbers. This variance has a pattern that indicates higher in deformation at N = 12, 16, 20 than at N = 10, 14, 18. Second is a high deformation, indicating almost a constant value with R ~ 3 above N = 20. Third is characterization associated with two-step phase transitions, from N = 18 through 20 to 22.

Following the characteristics $2^+$ and $4^+$ energies, it is found that two pseudo-shells are revealed along with one pseudo-shell within the space N = 8 and 20. Along the $2^+$ systematic path, we find a shell gap at N = 8 and N = 14, respectively. More importantly, by forcing the point at N = 20 towards a higher value close to the value at N = 14, as shown in Fig. 4, we can see another shell gap at N = 20. In contrast, the curve with the $4^+$ draws a pattern indicating a deformation centered at the pseudo-shell with N = 8 to 20. It is easily expected, following the systematic trend, that at N = 8 the point would be around between 2000 keV and 2500 keV, yielding an R value to be close to 1.7. See such an assumed point at N = 8 in Fig. 4. We address how to explain this controversial behavior between the $2^+$ and the $4^+$ excitations, revealing such a fluctuated deformation along neutron numbers. Interestingly, the systematic of the second $0^+$ looks a similar trend of the R values. It implies that there would be shape coexistence in the corresponding Mg nuclei. On the other hand, any similar aspects are not seen at Z = 10 or at Z = 14.

As far as we are concerned with systematic features in the space 8 $\leq$ Z $\leq$ 20, it is clear that a phase transition occurs at N = 20 out of N = 18. The phase transition at N = 20 is also observed in the Z = 10 system but not in the Z = 14. According to our previous work aiming at studying the ferro-deformation near the critical point of Z = 8, N = 20, the configuration Z = 12, N = 20 or 22 cannot yield a large deformation since it has the Hund configuration (2, 6) [11]. Furthermore, beyond N = 20 the deformation increases as saturating at R ~ 3.1. This phenomenon has been widely studied as being called the 'island of inversion', including the Ne and Al nuclides [7, 12-24].

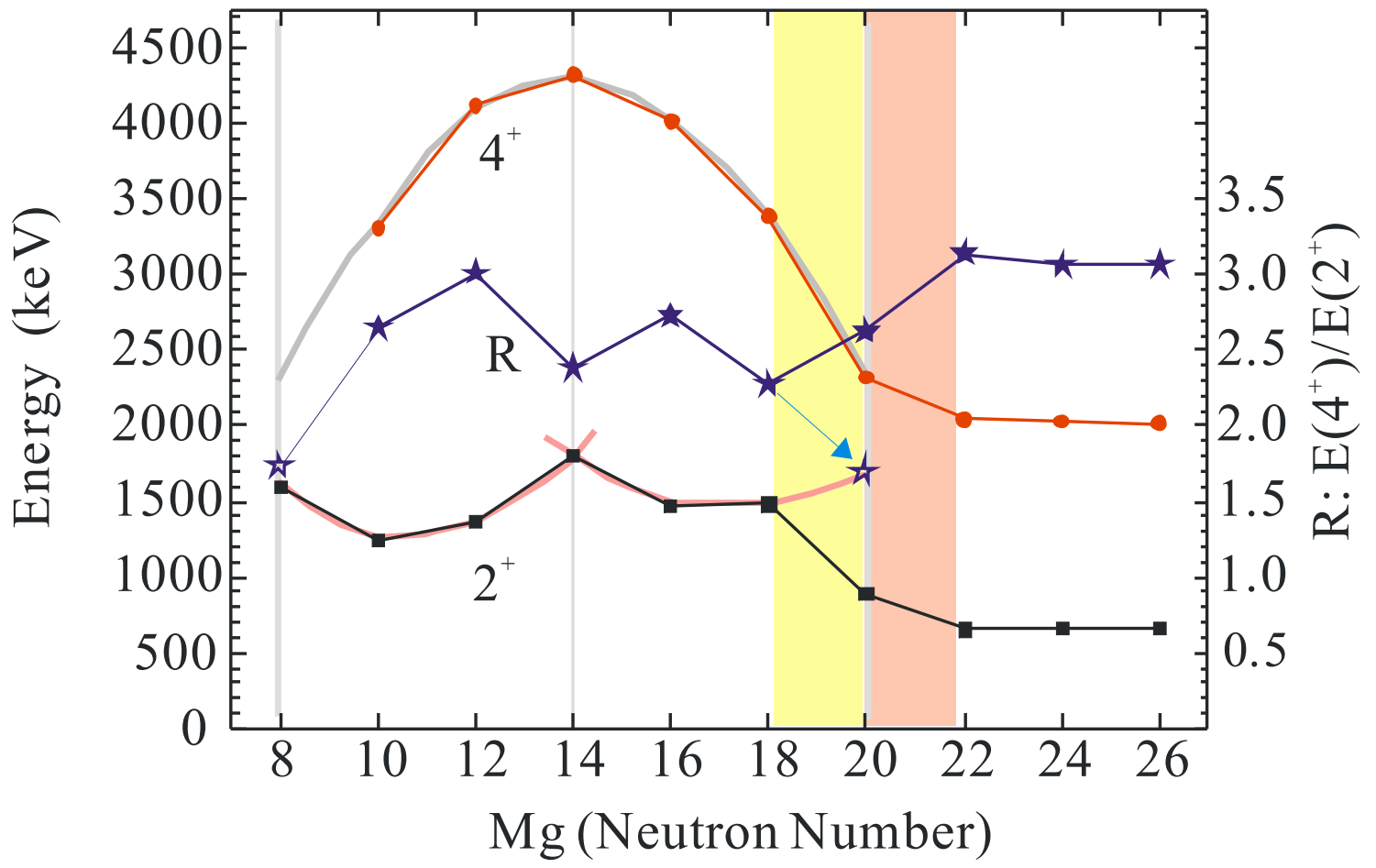

Fig. 4. The same as Fig. 3(b), but includes assumed points along with smooth lines for emphasizing shell gaps. Moreover, phase transition regions are denoted in square color boxes. Points at N = 8 and 20 are based on the assumed values.

Nevertheless, over the above characteristics, the utmost remarkable feature is the very high deformation at Z = 12, N = 12, $^{24}$Mg. On the basis of current shell model descriptions, it is hard to explain certainly such a high deformation. It is to require another approach to understanding of this feature, indicating a static large deformation in the ground state. By arguing that this high deformation would come from a geometrical symmetry as seen in molecules, we introduce an octahedral molecular structure comprising six points with α particles.

### 3. Octahedral symmetry: Molecular structures

As pointed out earlier, for understanding the unpredictable feature at the system with Z = 12 and N = 12, we shall start by developing a geometrical model based on molecular structures. Figure 5 describes the molecular structures for the α particle resonance nuclides; $^{24}$Mg, $^{20}$Ne, $^{16}$O, and $^{12}$C. It should be noted that the points in each structure correspond to α particles. Two distinctive features are seen. First, the nuclei $^{24}$Mg and $^{20}$Ne are expected to be a large deformation built on the octahedral and hexahedral shapes while for the $^{16}$O and $^{12}$C a spherical symmetry is rather favored. Second, though it is well known, the nucleus $^{8}$Be is not formed in stability with two α particles. Provided that the $^{8}$Be system consists of 4 protons and 4 neutrons for the ground stability, it is not possible to explain the very instability, leading to two α emissions, because 8 nucleons manage to construct a stable structure. From this observable, we can draw an important conclusion that α particles play a decisive role in controlling nuclear structures.

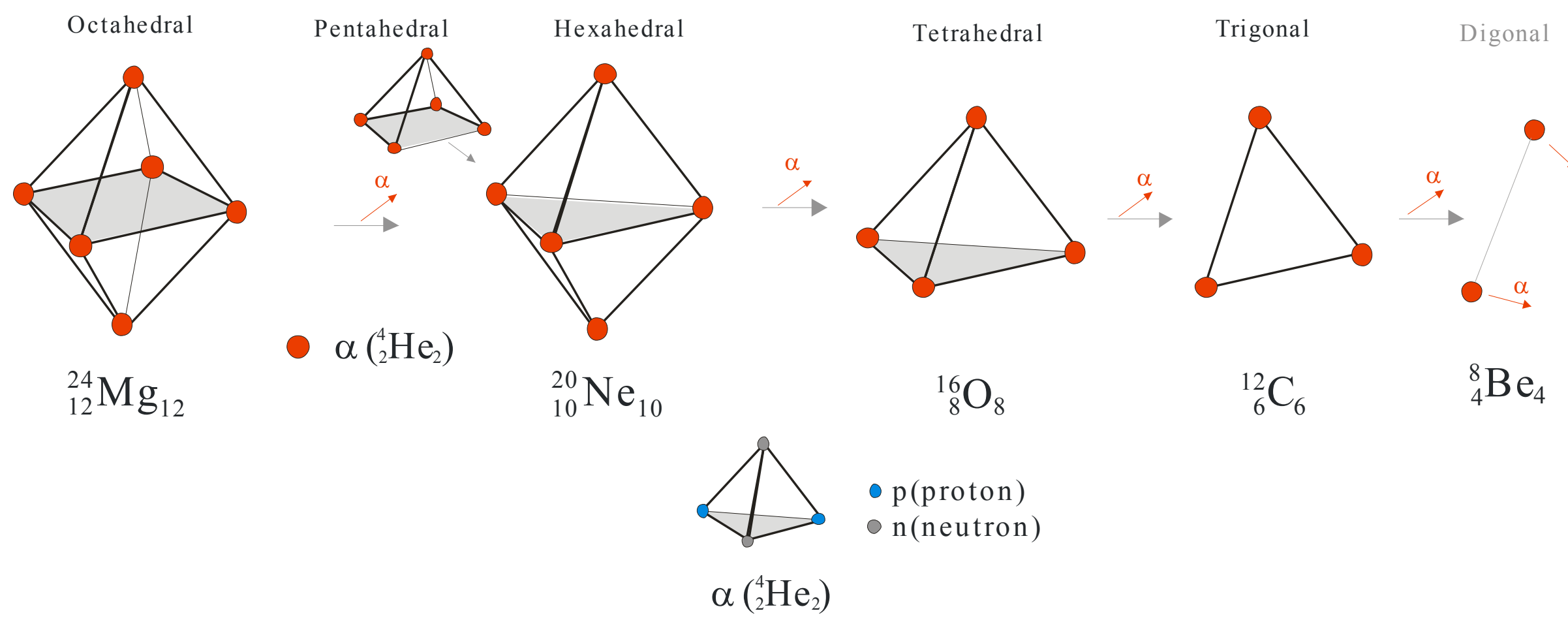

Fig. 5. Geometrical shapes correspond to the $^{24}$Mg, $^{20}$Ne, $^{16}$O, $^{12}$C nuclides, including the α geometry. The pentahedral (pyramidal) symmetry is converted into a hexahedron for more stabilizing. It is worthwhile to know that these nuclides are lying in a series of α decay, including the nucleus $^{8}$Be, which decays two α particles out of the ground state. For visualizing molecular structures, our models are described only with classical point of view.

According to the observation as shown in Fig. 5, α particles in a nucleus should be a basic constituent like a nucleon, protons or neutrons. We suggest that *the α particle should be a third nucleon, by naming it the 'Alpharon'*. With the results for the study of molecular structures of the He nuclides, we limit a bonding capacitance of an alpharon to four nucleons. However, in some circumstances, more nucleons may be allowed. In this case, neutrons distribute over an alpharons-skeleton as interacting with themselves, like π electrons in a molecule. We will discuss for more details later.

For the case of $^{24}$Mg, as shown in Fig. 5, it is fitted to forming an octahedron combined by six points with alpharons. It is easily expected that this octahedral structure leads to a large deformation by a symmetry breaking like the Jahn-Teller distortion in molecules [2]. We notice that, usually, the $^{24}$Mg has been studied with a cluster of the $^{20}$Ne with an α particle. Turning our attention to the Z = 10, N = 10 system, we know that a hexahedral shape would give rise to deformation in the ground state. Moreover, provided that two neutrons are positioned at top and down points as shown in Fig. 6, the nucleus $^{22}$Ne with N = 12 yields also a large deformation. See the systematic features regarding deformation of the Ne nuclides in Fig. 2(b).

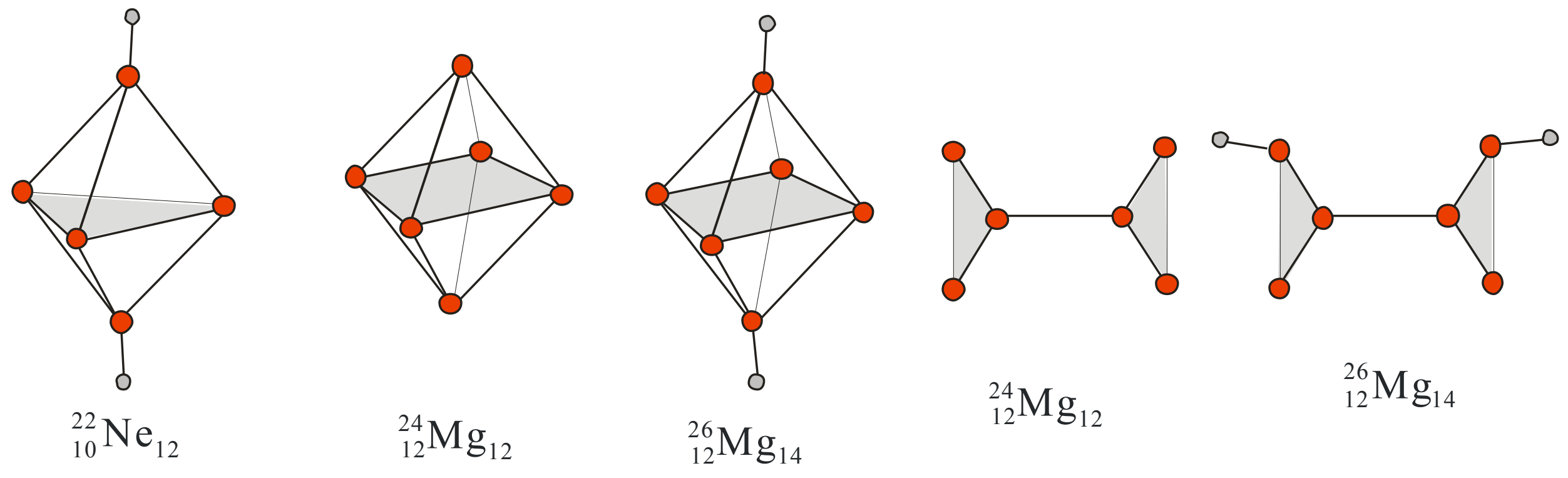

Fig. 6. Geometrical models corresponding to the $^{22}$Ne, $^{24}$Mg, and $^{26}$Mg nuclides. Notice that two alpharons at top and down points in $^{26}$Mg based on the octahedral geometry have five bonding branches, respectively.

If we go to heavier masses than $^{24}$Mg, however, the octahedral system is no longer valid for them. The reason, as already pointed out, comes from a restricted bonding capacitance of an alpharon. In order to overcome this problem, it is needed to replace an octahedron with another geometrical structure. By assuming that a molecular Ethene(Ethylene) skeleton is one of favored configurations for the Z = 12, N = 12 system, we introduce an Ethene-like structure as shown in Fig. 6, where both the octahedral and the ethene-like geometries are demonstrated. It should be again emphasized that the $^{22}$Ne system is still consistent with the hexahedral structure, leading directly to a large deformation.

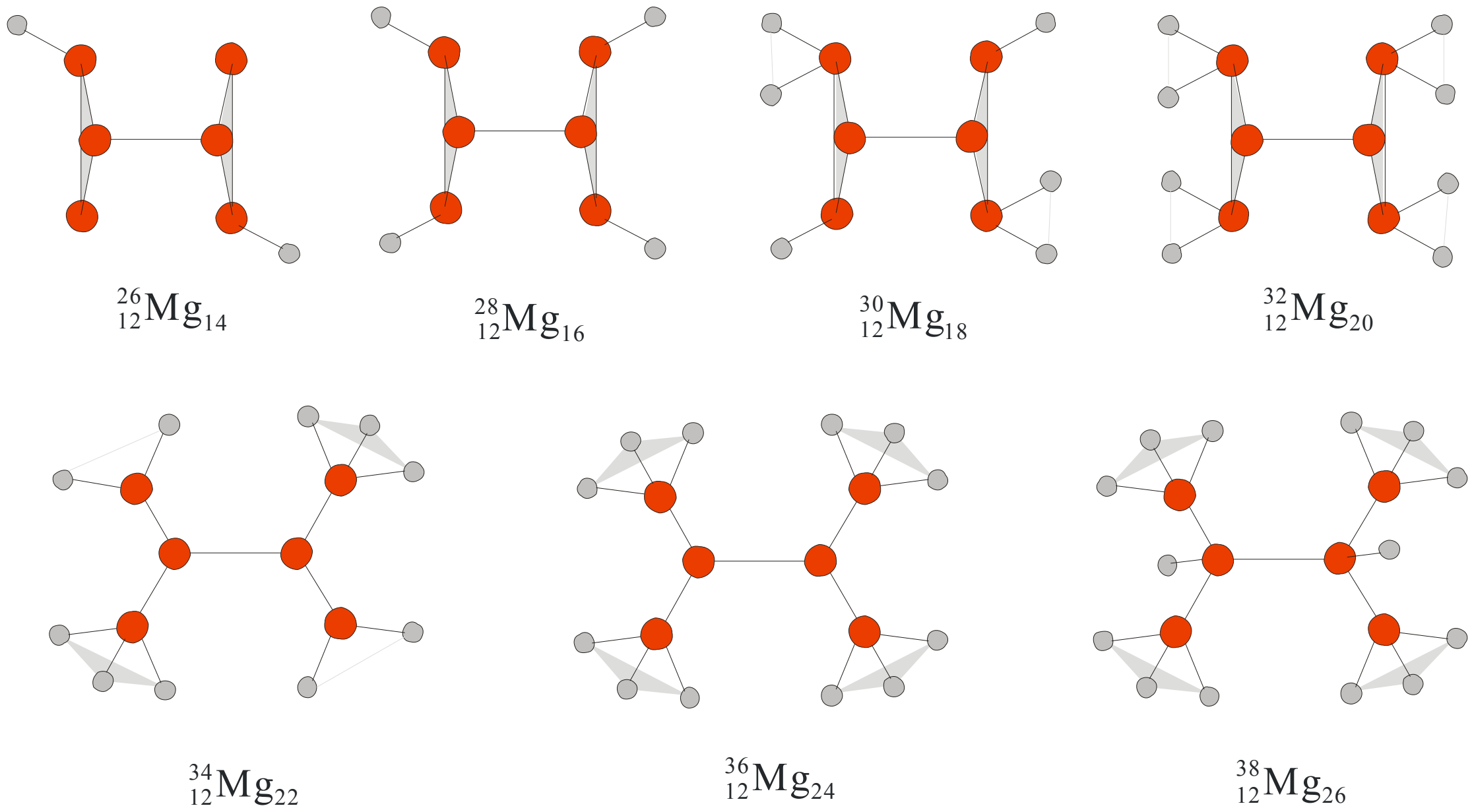

Fig. 7. Molecular structure descriptions for the Mg nuclides with N > 12. It should be noted that neutron bonding configuration would be in many ways, as being delocalized over the entire of alpharons skeleton.

As shown in Fig. 7, with this ethene-like structure model, we can extend the Z = 12 system to with N = 26, $^{38}$Mg. As adding more and more neutrons to four alpharon points, which are linked to two points in an equatorial line, we are able to construct the Mg nuclides with N > 12. Needless to say, the neutron bonding configuration would be in many ways, for instance, rather favoring a pair-like coupling. Density functional theory, with a basis of our model, may describe possible structures, including binding energies and neutron distributions.

At N = 20, each four points are fully occupied with two neutrons. It is worthwhile to know that three-alpharon points in left and right side with respect to the equatorial bonding axis, are connected with themselves, leading to a trigonal structure. At N = 22, however, such a trigonal bond has to be broken for adapting more neutrons. This breaking induces a more deformation. Now we understand, with this model, why the Mg nuclides with N = 22, 24, and 26 have the very deformation. Through N = 24, six alpharons complete their bonding at N = 26. In contrast, within the space N = 14 to N = 20, the fluctuation in deformation can be explained in terms of conditions of the symmetrical or the asymmetrical with respect to the equatorial center line.

We wish to put a remark that there would be a possibility of shape coexistence in $^{24}$Mg due to the octahedral on one hand and the ethene-like structures on the other hand. Moreover, it is highly interesting to know whether the $^{40}$Mg, with N = 28, system lies at the drip line or not. A possible scenario is a halo structure associated with two neutrons surrounding all over the core $^{38}$Mg. In this case, the halo-like distribution would make a large deformed orbital, which indicates the N = 28 shell gap breaking.

## 4. The third nucleon α particles: Alpharons

Figure 8 describes nuclear geometrical structures based on our hypothetical model theory for the He and Be nuclides, including, for comparisons, some molecules.

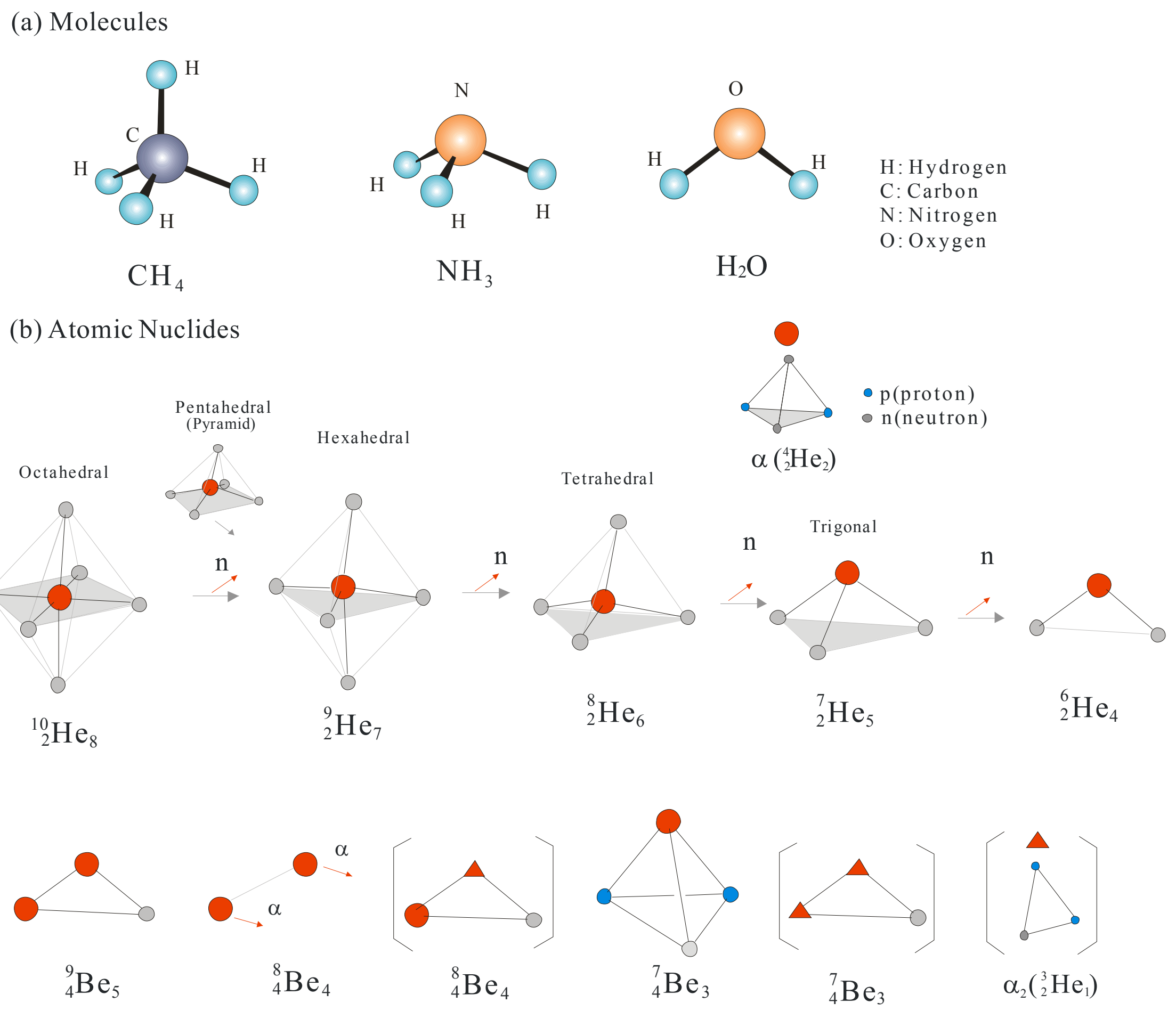

Fig. 8. Plots of molecular structures, (a) for the methane, ammonium, and water molecules, and (b) for the He and Be nuclides. The structures based on the $^{3}$He nucleons, $\alpha_2$, in square brackets turns out to be impossible in a nucleus. See the text for more details.

For structure descriptions of the nuclei with Z > N, we shall assume that the nucleus $^{3}$He would play a role like an Alpharon. Following this assumption, let us introduce another Alpharon, the nucleus $^{3}$He, and we call it *Alpharon2 by denoting* $\alpha_2$. See Fig. 8. For confirming our assumption, firstly, we focus on the nucleus $^{8}$Be that is known two-α emission activity. If we apply an $\alpha_2$ to making the structure of $^{8}$Be, it would be a trigonal shape, eventually, which leads to a more stable nucleus owing to three body interactions. This is not the case from the experimental observables. In turn, let us look at the nucleus $^{7}$Be. As shown in Fig. 8, two geometries are possible; one is a tetrahedral structure with an alpharon, two protons, and a neutron, and on the other hand there is a trigonal structure with two $\alpha_2$ and a neutron. By observing the very stability of $^{7}$Be, with a half-life of 53.24 days [6], we expect that the tetrahedral structure should prefer to the trigonal one. Furthermore, in proton-rich circumstances, nuclear structures have to be described in terms of shell model schemes based on excess protons. Under the $\alpha_2$ structural conditions, this is not the case since proton nucleons are supposed to be tightly bounded in the $\alpha_2$. Therefore, we draw a conclusion that *the alpharon, the nucleus* $^{4}$*He, is unique as a third nucleon inside a nucleus*. On the other hand, **the Z = 2, N =1 trigonal system, the nucleus $^{3}$He, should be an open system in a nucleus**.

We turn our attention to the structures of the He nuclides, as focusing on $^{10}$He and $^{8}$He. First of all, notice the very similarity between our model descriptions for the He nuclides and those for the molecules as shown in Fig. 1 and Fig.

8(a). According to the limited bonding capacitance, the nucleus $^{10}$He is not allowed for stability like $^{8}$He, with a half-life 119 ms [6]. Nevertheless, the ground state of $^{10}$He is formed as a resonance combined by the $^{8}$He and two neutrons, indicating $T_{1/2}$ ~ 300 keV [6]. We suggest that such a $^{10}$He structure should be built on an octahedral symmetry as shown in Fig. 8. It is well known that both $^{6}$He and $^{8}$He have neutron halos; two neutrons and four neutrons, respectively. According to our geometrical models, it is easily understood why they do have such neutron halos. The $^{10}$He structure would be viewed as having double halos; one is four neutrons over the $^{4}$He core and the other is two neutrons surrounding the $^{8}$He core.

Now we turn again to the Mg nuclides. We show an example of molecular structures for the $^{20}$Mg and $^{22}$Mg in Fig. 9. We wish to emphasize that the structure of $^{20}$Mg is based on a tetrahedral skeleton, while the $^{22}$Mg is based on a chain skeleton having two pyramidal sub-structures. Our model reveals an evidence of, on one hand, spherical structure for $^{20}$Mg, on the other, deformed structure for the case of $^{22}$Mg. Moreover, we notice that the skeleton of $^{20}$Mg is the nucleus $^{16}$O, of which isobaric nucleus corresponds to the $^{20}$O. It is very interesting to make a comparison between some characteristic states dominated by four protons or four neutrons in these isobaric nuclides. We suggest that a measurement should be made for identifying the corresponding asymmetric mixed states in $^{20}$O and $^{20}$Mg.

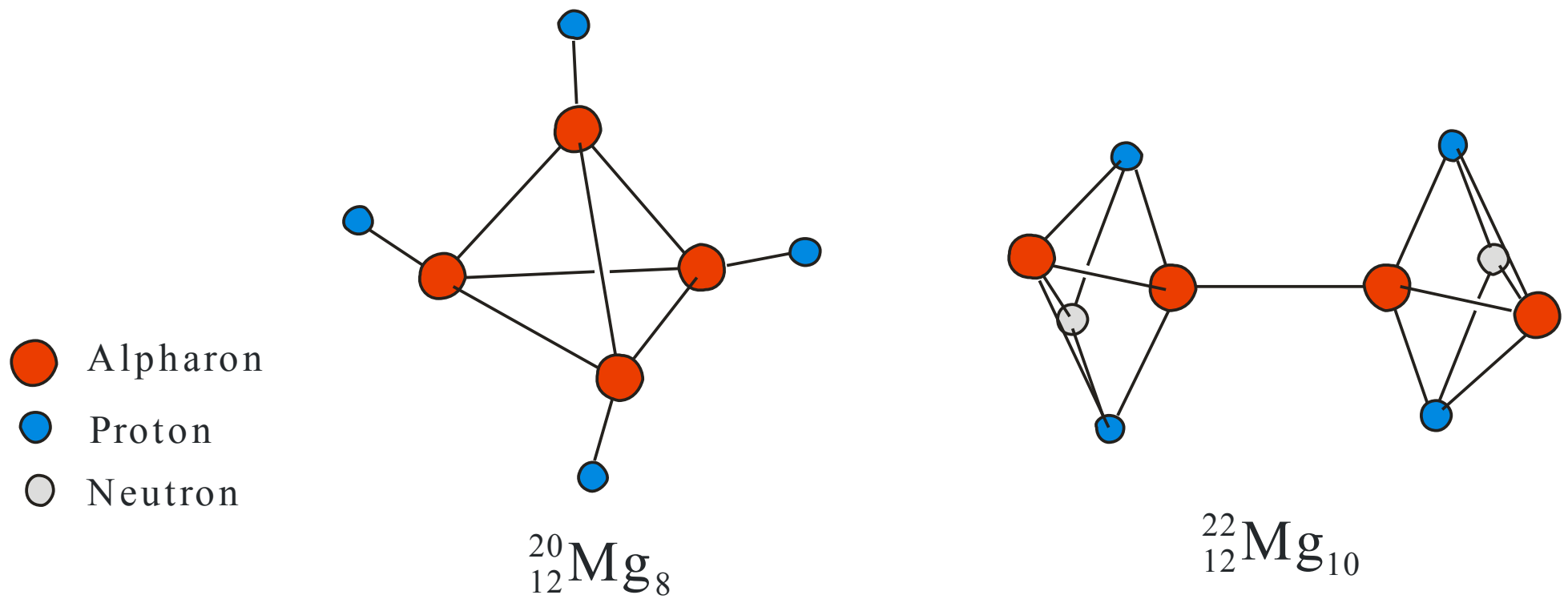

Fig. 9. Model descriptions of geometrical structures for the $^{20}$Mg and $^{22}$Mg nuclides.

Next, let us again examine collectivity of the Z = 12, N = 24 system along with the Z = 10, N = 20, with another point of view. If twelve (ten) neutrons are coupled with themselves as being bonded weakly to their alpharon cores, the neutrons might distribute coherently over the Z = 12, N = 12 (Z = 10, N = 10) core. One of the most important aspects of this model is a whole distribution of neutrons over the surface. In this case, we expect that there would be a possibility for revealing a skin structure. The density functional calculations may provide useful mapping representations in this regard. We point out that the deformation of the Ne nuclides with N = 20 and 22 could be interpreted in terms of the ferro-deformation concept as well [11].

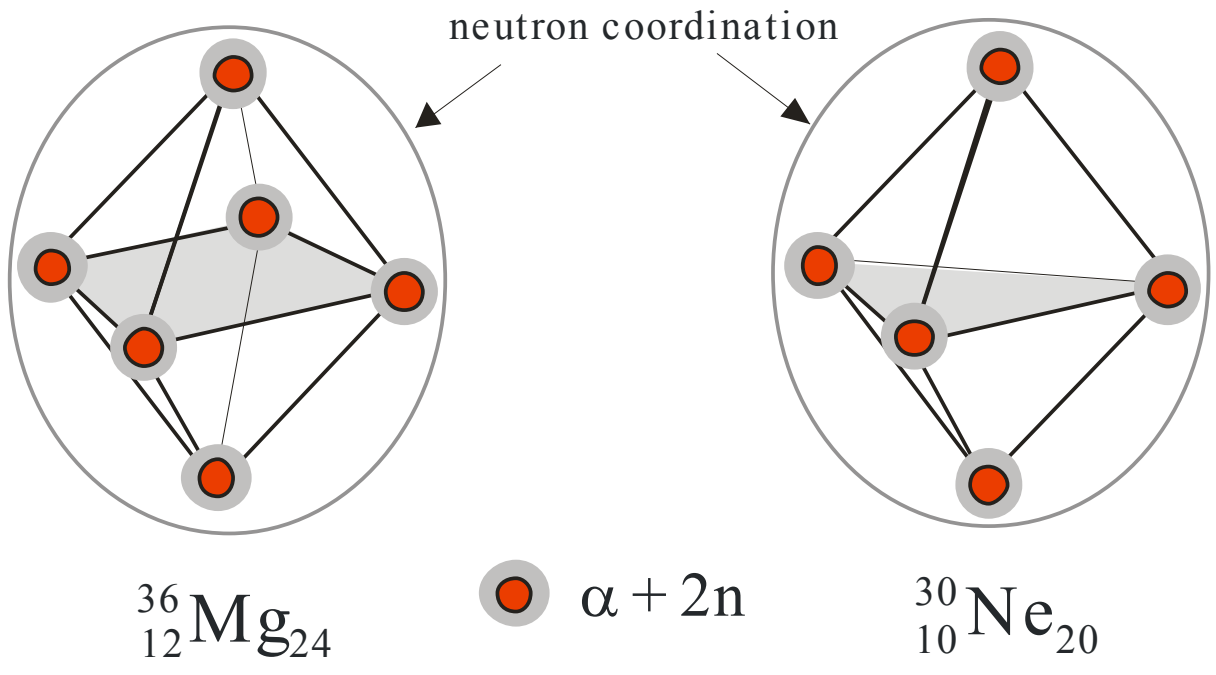

Fig. 10. A model for the $^{36}$Mg and $^{30}$Ne nuclei. Neutrons distribute like a skin as denoted by neutron coordination all over the core of $^{24}$Mg and the core of $^{20}$Ne, respectively.

## 5. Conclusions

We presented that the emergence of a large deformation in Mg nuclides, especially at the Z = 12, N = 12, should be associated with an octahedral molecular structure. Within the framework of geometrical symmetry, we found that the Z = 12, N = 12 system would organize the octahedron consisting of six points of $\alpha(^{4}He)$ particles. With this point of view, we were able to conclude such that; the Z = 10, N = 10, $^{20}Ne$, corresponds to a hexahedral, the Z = 8, N = 8, $^{16}O$, does to a tetrahedral, the Z = 6, N = 6, $^{12}C$, does to a trigonal symmetries. The Z = 2, N = 2 system, $^{4}He(\alpha)$, fits a tetrahedral symmetry with four points of nucleons; two protons and two neutrons. The enhanced deformation at Z = 12 with N > 20 could be explained by a deformed distribution of surrounding neutrons over the $^{24}Mg$ ethene-like skeleton. We suggested that the α particle should be treated as a third nucleon and we defined its name to be the '***Alpharon***'. In contrast, the Z = 2, N = 1 trigonal symmetry, the nucleus $^{3}He$, is found to be an open system in a nucleus. We addressed a question about the existence of collective structures associated with both an octahedral and an ethene-like geometry in $^{24}Mg$. In addition, we proposed that a measurement should be made for identifying the asymmetric mixed states in $^{20}O$ and $^{20}Mg$.

Our model offers new insights into understanding of nuclear many-body systems; shell structures including collectivity, nucleon-synthesis including α, β decays, and nuclear matters. We hope that the present model, driven by the search for understanding patterns, might open a new gate on the field of nuclear physics. We conclude that nature favors three-body systems rather than two-body systems in the quantum world; three quarks for a nucleon, three nucleons for a nucleus. Finally we address what is the nature of nucleon-nucleon interaction in a nucleus?

## References

[1] Peter Atkins and Julio de Paula, *Physical Chemistry*, Chapter 11 (Oxford, 2010).
[2] M. C. M. O'brien and C. C. Chancey, Am. J. Phys. **61**, 688 (1993).
[3] J. R. Lakowicz, *Principles of Fluorescence Spectroscopy* (Springer, New York, 2006).
[4] J. Dudek, A. Gozdz, and N. Schunck, **arXiv**: 0303001 (2003).
[5] J. Dudek, D. Curien, N. Dubray, J. Dobaczewski, V. Pangon, P. Olbratowski, and N. Schunck, Phys. Rev. Lett. **97**, 072501 (2006).
[6] National Nuclear Data Center, Brookhaven National Laboratory, http:// www.nndc.bnl.gov/ (February 2016).
[7] P. Doornenbal, H. Scheit, S. Takeuchi , N. Aoi, K. Li, M. Matsushita, D. Steppenbeck, H.Wang, H. Baba, H. Crawford, C. R. Hoffman, R. Hughes, E. Ideguchi, N. Kobayashi, Y. Kondo, J. Lee, S. Michimasa, T. Motobayashi, H. Sakurai, M. Takechi, Y. Togano, R. Winkler, and K. Yoneda, Phys. Rev. Lett. **111**, 212502 (2013).
[8] Chang-Bum Moon, **arXiv**:1604.01017 (2016).
[9] Chang-Bum Moon, **arXiv**:1604.02786 (2016).
[10] Chang-Bum Moon, **arXiv**:1604.05013 (2016).
[11] Chang-Bum Moon, **arXiv**:1605.00370 (2016).
[12] E. K. Warburton, J. A. Becker, and B. A. Brown, Phys. Rev. C **41**, 1147 (1990).
[13] R. Klapisch, C. Thibault-Philippe, C. De´traz, J. Chaumont, R. Bernas, and E. Beck, Phys. Rev. Lett**.** **23**, 652 (1969).
[14] C. Thibault, R. Klapisch, C. Rigaud, A. Poskanzer, R. Prieels, L. Lessard, and W. Reisdorf, Phys. Rev. C **12**, 644 (1975).
[15] X. Campi, H. Flocard, A. Kerman, and S. Koonin, Nucl. Phys. A **251**, 193 (1975).
[16] A. Poves and J. Retamosa, Phys. Lett. B **184**, 311 (1987).
[17] K. Yoneda et al., Phys. Lett. B **499**, 233 (2001).
[18] A. Gade, P. Adrich, D. Bazin, M. D. Bowen, B. A. Brown, C. M. Campbell, J. M. Cook, S. Ettenauer, T. Glasmacher, K. W. Kemper, S. McDaniel, A. Obertelli, T. Otsuka, A. Ratkiewicz, K. Siwek, J. R. Terry, J. A. Tostevin, Y. Utsuno, and D. Weisshaar, Phys. Rev. Lett. **99**, 072502 (2007).
[19] S. Takeuchi, N. Aoi, T. Motobayashi, S. Ota, E. Takeshita, H. Suzuki, H. Baba, T. Fukui, Y. Hashimoto, K. Ieki, N. Imai, H. Iwasaki, S. Kanno, Y. Kondo, T. Kubo, K. Kurita, T. Minemura, T. Nakabayashi, T. Nakamura, T. Okumura, T. K. Onishi, H. Sakurai, S. Shimoura, R. Sugou, D. Suzuki, M. K. Suzuki, M. Takashina, M. Tamaki, K. Tanaka, Y. Togano, and K. Yamada, Phys. Rev. C **79**, 054319 (2009).
[20] P. Doornenbal, H. Scheit, N. Kobayashi, N. Aoi, S. Takeuchi, K. Li, E. Takeshita, Y. Togano, H. Wang, S. Deguchi, Y. Kawada, Y. Kondo, T. Motobayashi, T. Nakamura, Y. Satou, K. N. Tanaka, and H. Sakurai, Phys. Rev. C **81**,

041305(R) (2010).
[21] P. Doornenbal, H. Scheit, S. Takeuchi, N. Aoi, K. Li, M. Matsushita, D. Steppenbeck, H. Wang, H. Baba, E. Ideguchi, N. Kobayashi , Y. Kondo, J. Lee, S. Michimasa, T. Motobayashi , A. Poves, H. Sakurai, M. Takechi, Y. Togano , and K. Yoneda, Phys. Rev. C **93**, 044306 (2016).
[22] P. Doornenbal, H. Scheit, N. Aoi, S. Takeuchi, K. Li, E. Takeshita, H. Wang, H. Baba, S. Deguchi, N. Fukuda, H. Geissel, R. Gernhauser, J. Gibelin, I. Hachiuma, Y. Hara, C. Hinke, N. Inabe, K. Itahashi, S. Itoh, D. Kameda, S. Kanno, Y. Kawada, N. Kobayashi, Y. Kondo, R. Krucken, T. Kubo, T. Kuboki, K. Kusaka, M. Lantz, S. Michimasa, T. Motobayashi, T. Nakamura, T. Nakao, K. Namihira, S. Nishimura, T. Ohnishi, M. Ohtake, N. A. Orr, H. Otsu, K. Ozeki, Y. Satou, S. Shimoura, T. Sumikama, M. Takechi, H. Takeda, K. N. Tanaka, K. Tanaka, Y. Togano, M. Winkler, Y. Yanagisawa, K. Yoneda, A. Yoshida, K. Yoshida, and H. Sakurai, Phys. Rev. Lett. **103**, 032501 (2009).
[23] P. Doornenbal, H. Scheit, S. Takeuchi, Y. Utsuno, N. Aoi, K. Li, M. Matsushita, D. Steppenbeck, H. Wang, H. Baba, E. Ideguchi, N. Kobayashi, Y. Kondo, J. Lee, S. Michimasa, T. Motobayashi, T. Otsuka, H. Sakurai, M. Takechi, Y. Togano, and K. Yoneda, Prog. Theor. Exp. Phys. 053D01 (2004).
[24] E. Caurier, F. Nowacki, and A. Poves, Phys. Rev. C **90**, 014302 (2014).